\documentclass[
reprint,
superscriptaddress,
amsmath,amssymb,
aps,prx,
longbibliography]{revtex4-2}

\usepackage{amsfonts}

\usepackage{graphicx}
\usepackage{dcolumn}
\usepackage{bm}
\usepackage{hyperref}
\usepackage{physics}
\usepackage{xcolor}
\usepackage{mathrsfs}
\usepackage{mathtools}
\usepackage{comment}

\begin{document}

\title{Stabilization of Discrete Time-Crystaline Response on a Superconducting Quantum Computer by increasing the Interaction Range}

\author{Andrea Solfanelli}
 \affiliation{SISSA, via Bonomea 265, I-34136 Trieste, Italy}
 \affiliation{INFN, Sezione di Trieste, Via Valerio 2, 34127 Trieste, Italy}
 \affiliation{Center for Life Nano-Neuro Science @ La Sapienza, Italian Institute of Technology, 00161 Roma, Italy}
\email{asolfane@sissa.it}

\author{Stefano Ruffo}
\affiliation{SISSA, via Bonomea 265, I-34136 Trieste, Italy}
 \affiliation{INFN, Sezione di Trieste, Via Valerio 2, 34127 Trieste, Italy}
\affiliation{Istituto dei Sistemi Complessi, Via Madonna del Piano 10, I-50019 Sesto Fiorentino, Italy}

\author{Sauro Succi}
\affiliation{Center for Life Nano-Neuro Science @ La Sapienza, Italian Institute of Technology, 00161 Roma, Italy}
\affiliation{Physics Department, Harvard University, Oxford Street 17, Cambridge, USA}

\author{Nicol\`o Defenu}
\affiliation{Institut f\"ur Theoretische Physik, ETH Z\"urich, Wolfgang-Pauli-Str. 27 Z\"urich, Switzerland}
\email{ndefenu@phys.ethz.ch}
\date{\today}

\begin{abstract}
The simulation of complex quantum many-body systems is a promising short-term goal of noisy 
intermediate-scale quantum (NISQ) devices. However, the limited connectivity of native qubits 
hinders the implementation of quantum algorithms that require long-range interactions. 
We present the outcomes of a digital quantum simulation, where we overcome the limitations
of the qubit connectivity in NISQ devices. Utilizing the universality
of quantum processor native gates, we demonstrate how to implement couplings among physically disconnected
qubits at the cost of increasing the circuit depth. 
We apply this method to simulate a Floquet-driven quantum spin chain featuring 
interactions beyond nearest neighbors. Specifically, we benchmark the prethermal stabilization of 
the discrete Floquet time-crystalline response as the interaction range increases, a phenomenon 
never observed experimentally. 
Our quantum simulation addresses one of the significant limitations of superconducting quantum processors,
namely, device connectivity. It reveals that non-trivial physics involving couplings beyond nearest
neighbors can be extracted after the impact of noise is properly taken into account in the theoretical
model and consequently mitigated from the experimental data.
\end{abstract}

\maketitle
\section{Introduction}
Quantum computers are a revolutionary technology that have the potential to transform our society by solving problems that classical computers cannot \cite{Preskill18QUANTUM2}. However, these machines are still subject to uncontrollable noise and errors that limit their performance, which are far from the threshold required for error correction. Despite these limitations, recent progress in the realm of noisy intermediate-scale quantum (NISQ) devices represents an exciting opportunity for many-body physics by introducing new laboratory platforms with unprecedented control and measurement capabilities\cite{Ippoliti2021PRXQuantum}. Quantum simulation of the dynamics of more and more complex quantum many-body systems is expected to be one of the most promising short-term goals of NISQ quantum computing devices, with intriguing applications in diverse areas ranging from quantum chemistry \cite{Kassal2011AnnRevPhysChem,Hastings2015QuantumInfoComput,Cao2019ChemRev} and material science \cite{DeLeon2021Science} to high-energy physics \cite{Nachman2021PRL}. 
		
		Various experimental platforms have been tested for quantum computing, among the others we can cite: trapped ions,\cite{CiracPRL1995,BenhelmNatPhys2008,LanyonScience2011,Haffner2008155} neutral Rydberg atoms \cite{Endres2016Science,Graham2019PRL,Scholl2021Nature}, coherent photons \cite{Broome2013Science,Zhong2021PRL}, nuclear spins in molecule \cite{Cory1998PRL,Vandersypen2001Nature}, NV centers \cite{Cappellaro2009PRL,Randall2021Science} and superconducting qubits \cite{Devoret2013Science,BlaisRevModPhys}. Each of them has its own advantages and drawbacks \cite{Cheng2023}. In this paper, we focus on superconducting quantum processors. Superconducting qubits are relatively easy to fabricate and can be densely packed, allowing for the construction of large-scale quantum computers. This makes them a promising platform for scaling up quantum computing applications \cite{Wendin2017RepProgPhys}. Moreover they can be manipulated with a wide range of microwave frequencies, making them versatile and flexible for implementing various quantum gates \cite{Devoret2013Science}. 
		
Thanks to this flexibility, the number of quantum simulations implemented on noisy superconducting devices has steadily risen in recent years, also thanks to the possibility to easily access these machines remotely, allowing to benchmark a number of phenomena which were not or were very little experimentally corroborated before. Among this plethora of studies we may mention the following: the observation of disorder stabilized discrete time crystal phases,\cite{Mi2022Nature,Frey2022SciAdv} the realization of topologically ordered states, dynamical topological phases and topological edge states \cite{Satzinger2021Science,Zhang2022Nature,Mi2022Science}. One should also cite the observation of Leggett-Garg's inequalities violations \cite{Santini2022PRA}, the validation of dynamical scalings \cite{Quintero2022FQST,Keenan2023npjQI} and several studies in the context of quantum thermodynamics \cite{Solfanelli2021PRXQ,Solfanelli2022AVSQ}.
On the other hand, the performance of superconducting NISQ devices is limited by the presence of various sources of noise and decoherence, whose impact grows with the depth and complexity of the quantum circuit realized, limiting the investigation of non-local effects and complex geometries. 
		
Long-range interactions are known to boost the performance of quantum hardware \cite{MonroeRevModPhys2021,ChomazArXiv2022,periwal2021programmable, solfanelli2023quantumheatengine} as they evade the traditional constraint imposed by thermal equilibration and noise propagation. The stability of long-range quantum systems against external perturbations \cite{xuPhysics2022} and their role as a source of unprecedented phenomena, including novel forms of dynamical phase transitions and defect formation \cite{defenu2019dynamical,DefenuPRL2018,DefenuPRB2019,HalimehPRR2020,Halimeh2017Prethermalization}, anomalous thermalization and information spreading \cite{RegemortelPRA2016,TranPRX2020,ChenPRL2019,KuwaharaPRX2020}, metastable phases \cite{DefenuPNAS2021,giachetti2023entanglement}, and entanglement scalings \cite{Ares2018PRA, solfanelli2023logarithmic,Lakkaraju2022Mimicking} have been widely proven \cite{defenu2023long,xuPhysics2022}. However, the theoretical comprehension of such behaviors is still mainly limited to integrable quadratic systems or perturbations of fully connected mean-field models, while systems with a tunable interaction range require an extremely high degree of experimental control \cite{defenu2023long}.
		
In this context, one main limitation of superconducting NISQ devices is their extremely limited connectivity, since superconducting qubits are typically arranged in a one or two-dimensional grid with nearest-neighbor connectivity, making it challenging to implement quantum algorithms that require long-range interactions \cite{Preskill18QUANTUM2}. In this paper, we aim to advance the field of digital quantum simulation on superconducting quantum hardware by investigating the possibility of reproducing the dynamics of systems with couplings beyond nearest neighbors. To achieve this, we utilize the universality of the quantum processor native gates to implement couplings among physically unconnected qubits. While the depth of the resulting quantum circuit increases with the effective range of the interaction, we show that careful consideration of gate noise, measurement errors, and statistical errors enables the removal of their effects from the raw results. The resulting error-mitigated data closely reproduce the theoretical expectations.
		
		More specifically, we implement the quantum simulation of a Floquet-driven quantum spin chain featuring interactions beyond nearest neighbors on IBM quantum superconducting processors. Indeed, the quantum circuit structure utilized by IBM quantum computers is well suited for implementing discrete Floquet driving protocols \cite{Ippoliti2021PRXQuantum}, making it a natural choice for such applications \cite{Frey2022SciAdv}. 
		
		Our focus is on the stabilization of discrete Floquet time-crystalline response as the interaction range increases. Discrete Floquet time crystals (DFTCs) are nonequilibrium many-body phases of matter that display a novel form of spatiotemporal order. In particular, in such phases the discrete time translation symmetry of the Floquet driving is broken and an order parameter exhibits persistent oscillations with a period which is an integer multiple of the period of the drive \cite{WilczekPRL2012,SachaPRA2015,ElsePRL2016,KhemaniPRL2016,BordiaNaturePhys2017,Zhang2017Nature,Choi2017Nature,Rovny2018PRL,Khemani2019arXiv,ElseAnnRevCondMattPhys2020}. The possibility of generating a DFTC in clean systems has been studied in the context of long-range interacting models, and our quantum simulation on IBM quantum processors constitutes its first experimental benchmark. Our results demonstrate the potential of superconducting quantum computing platforms to simulate quantum systems featuring interaction ranges going beyond the limits imposed by hardware connectivity and offer insights into the fundamental physics of long-range systems.
  \section{The Model}
  The kicked Ising spin chain is a prototypical model for the investigation of Floquet driven quantum systems, widely studied from a theoretical point of view \cite{RussomannoPRB2017,LerosePRB2019,PizziNatComm2021,Collura2022PhysRevX,GiachettiArXiv2022}. In this paper, we consider a driven quantum spin chain described by a time-dependent Hamiltonian of the form
	\begin{align}
		H(t) = -\sum_{j = 1}^{N}\sum_{r=1}^{R}J_rZ_jZ_{j+r}+ h(t)\sum_{j=0}^{N-1}X_j,\label{eq: kickedIsing}
	\end{align}
	where the time dependence is generated by a time periodic driving with period $T$ of the transverse magnetic field $h(t)$. The driving takes the form
	\begin{align}
		h(t) = \phi\sum_{n=1}^\infty\delta(t-nT).
	\end{align}
	The effect of this impulsive magnetic field applied at integer multiples of the driving period $t = nT$ is to impose a global rotation of every spin by an angle $2\phi$ along the $x$-axis. Accordingly, the Floquet dynamics is obtained by periodically intertwining the evolution generated by the Ising Hamiltonian at zero transverse field
	\begin{align}
		V = \prod_{j=1}^N\prod_{r = 1}^Re^{iTJ_r Z_jZ_{j+r}},
	\end{align}
	with the instantaneous kick operator,
	\begin{align}
		K_\phi = \prod_{j = 1}^{N}e^{-i\phi X_j}.
	\end{align}
	The resulting evolution operator for a single step of the Floquet protocol reads
	\begin{align}
		U_F = K_\phi V. \label{eq: UF}
	\end{align}
	The system is initialized at $t = 0$ in the fully polarized state with positive magnetization along the $\hat{z}$ direction $|\psi(0)\rangle = |\dots\uparrow\uparrow\uparrow\dots\rangle = |\dots0 0 0\dots\rangle$,
	where $|\uparrow\rangle$ and $|\downarrow\rangle$ denote the eigenstates of the $Z$ Pauli matrix with eigenvalues $+1$ and $-1$ respectively. In our case, these eigenstates correspond to the computational basis of the quantum processor, with the convention $|\uparrow\rangle = |0\rangle$ and $|\downarrow\rangle = |1\rangle$.
	
	The simplest realization of the time-crystalline spatiotemporal order is obtained by taking the kick operator $K_\phi$ to rotate each spin by an angle $\pi$ around a transverse axis $\hat{x}$. In this case, the kick operator is given by
	\begin{align}
		K_{\pi/2} = \prod_{j = 1}^{N}e^{-i\frac{\pi}{2} X_j} = \prod_{j = 1}^{N} X_j.
	\end{align}
	As a result, the time-evolved state after $n$ kicks, $|\psi(n)\rangle = U_F^n|\psi(0)\rangle$, exhibits a sequence of perfect jumps between the $|\dots\uparrow\uparrow\uparrow\dots\rangle$ and $|\dots\downarrow\downarrow\downarrow\dots\rangle$ states, leading to a persistent non-vanishing value of the order parameter in both space and time. The order parameter is given by
	\begin{align}
		m(n) = \langle\psi(n)|Z_j|\psi(n)\rangle = (-1)^n.
	\end{align}
	This is the simplest example of a subharmonic response, where the period of the order parameter evolution is twice the period of the Floquet driving. However, this behavior depends on the finely tuned choice of the kick angle $2\phi = \pi$. To observe a proper discrete time-crystalline phase of matter, the spatiotemporal order must be stable to sufficiently weak perturbations of the Hamiltonian parameters $\phi = \pi/2+\epsilon$, in the thermodynamic limit $N\to\infty$. This condition is generally not satisfied as the presence of the external driving leads to the exponential decay of the magnetization, ruling out long-lived oscillations. Protecting ordering against relaxation requires a mechanism to control the impact of dynamically generated excitations \cite{ElseAnnRevCondMattPhys2020}.
	
	In clean systems, the possibility of generating a DFTC has been studied in the context of long-range interacting models \cite{RussomannoPRB2017,SuracePRB2019,LerosePRB2019,PizziNatComm2021,Collura2022PhysRevX,GiachettiArXiv2022}, where the interaction between different lattice sites decays as a power law. However, for any finite $R$ and in the absence of disorder, the system magnetization exhibits an exponential decay with the number of Floquet steps:
	\begin{equation}
		m(n)\propto (-1)^ne^{-n\gamma_{\epsilon,R}}.
	\end{equation}
	 The decay rate $\gamma_{\epsilon,R}$ goes to zero as the perfect kick case is approached, i.e., for $\epsilon\to 0$. Moreover, as shown in Ref. \cite{Collura2022PhysRevX}, $\gamma_{\epsilon,R}$ is deeply affected by the interaction range. In the small $\epsilon$ limit, we have that
	\begin{align}
    \gamma_{\epsilon,R}\approx \epsilon^{2R+1}.
	\end{align}
	Therefore, increasing the interaction range exponentially enhances the order parameter lifetime. This difference in decay rate should already be apparent when comparing the nearest neighbor $R = 1$ and the next-to-nearest neighbor $R = 2$ cases. One of the main results of our digital quantum simulation is to demonstrate this increase in the order parameter lifetime.
 
\section{Quantum circuit implementation of the Floquet dynamics}\label{app: Quantum circuit implementation of the Floquet dynamics}
\begin{figure*}
		\centering
		\includegraphics[width=\linewidth]{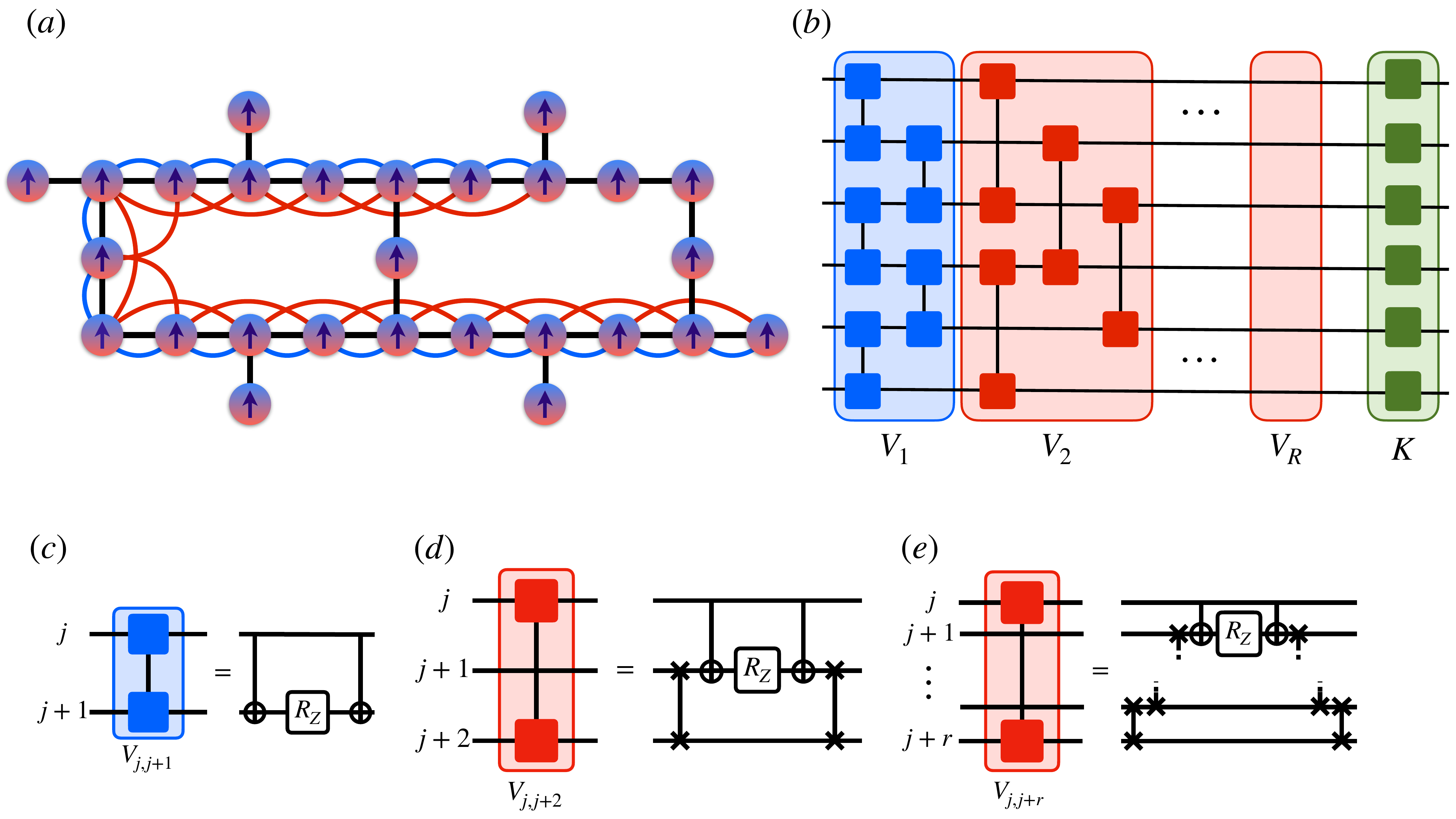}
		\caption{Quantum circuit implementation of Floquet dynamics with varying interaction range. (a) Topology of the \textit{ibmq\_mumbai} quantum processor. Black links represents the physical connections among the qubits on the quantum hardware, blue and red links represent the nearest neighbors $R = 1$ and next-to-nearest neighbor $R = 2$ Ising interactions we effectively implemented among physically unconnected qubits during our quantum simulation. (b) Quantum circuit implementing a single Floquet step for a kicked Ising model with $R$ range interactions. (c) Quantum gate implementation of nearest-neighbor Ising interaction among qubit $j$ and $j+1$. (d) Quantum gate implementation of next-to-nearest neighbor Ising interaction among qubit $j$ and $j+2$. (e) Quantum gate implementation of $r$-neighbor Ising interaction among qubit $j$ and $j+r$.}
		\label{fig: Quantum_circuit}
	\end{figure*}
The aim of our quantum simulation is to implement the dynamics of the Floquet-driven quantum spin chain with tunable interaction range, introduced in the previous section, on an IBM quantum processor. Specifically, we utilize the \textit{ibmq\_mumbai} 27-qubit processor, whose topology is depicted in Fig.\,\ref{fig: Quantum_circuit}a (further technical details can be found in the Supplemental Material (SM) Appendix). Our quantum circuit is optimized using the available connectivity and native gates of the device, including the controlled-$\mathrm{NOT}$ gate ($\mathrm{CNOT}$), the identity gate $\mathrm{ID}$, rotations along the z axis $R_Z$, the $\mathrm{NOT}$ gate $X$, and the $\mathrm{SX} = \sqrt{X}$ gate (see Appendix \ref{app:Quantum circuit optimization}). 

We notice that the Floquet unitary evolution operator at stroboscopic times $t=nT$ can be obtained by applying the unitary operator corresponding to each Floquet step $U_F$ $n$ times, i.e., $U(nT) = (U_F)^n$. Importantly, no Trotter approximation is required, which is a significant advantage of Floquet drivings, making them well-suited for quantum circuit implementation \cite{Ippoliti2021PRXQuantum}. Furthermore, the kicked Floquet protocol of interest can be further decomposed into the successive application of the kick operator $K_\phi$ and the Ising evolution operator $V$ (see Eq. \eqref{eq: UF}). The former can be expressed in terms of single-qubit gates, corresponding to local rotations along the $x$-axis, and the latter can be written as a product of mutually commuting unitaries that connect pairs of qubits at progressively larger distances as the interaction range is increased, i.e., $K_\phi = \prod_{i=1}^NR_{X,i}(2\phi)$ and $V = \prod_{r = 1}^R V_r$, respectively. Each $V_r$ can be implemented by applying the general method to effectively realize $r$-range interactions introduced in the previous section.
	
The quantum circuit corresponding to a single Floquet step is shown in Fig. \ref{fig: Quantum_circuit}(b), where blue gates represent nearest-neighbor Ising interactions $V_{j,j+1}$, red gates represent Ising interactions beyond nearest neighbors $V_{j,j+r}$, and green gates represent the final kick rotation $K_\phi$ applied equally to each qubit.
In particular, as shown in Fig. \ref{fig: Quantum_circuit}(c), the unitary operator associated to nearest-neighbor Ising interactions can be decomposed in terms of the elementary gates as
	\begin{align}
		V_{j,j+1} &= e^{iTJ_1 Z_jZ_{j+1}}\notag
		\\
		&= \mathrm{CNOT}_{j,j+1}R_Z(2J_1T)\mathrm{CNOT}_{j,j+1}.\label{eq: nn Ising}
	\end{align}
On the other hand, the limited processor connectivity does not allow for a simple decomposition of $r$-range Ising interactions. The idea to overcome this limitation is to exchange the qubit states by applying a sequence of $\mathrm{SWAP}$ gates among the couples of physically connected qubits that lie between $j$ and $j+r$. By doing so, the initial state of qubit $q_{j+r}$ is effectively encoded in qubit $q_{j+1}$. Specifically, we achieve this by applying the gate sequence
\begin{align}
	S_r = \prod_{l=1}^{r-1}\mathrm{SWAP}_{j+l,j+l+1}.\label{eq: SWAPs}
\end{align}
Next, we apply $V_{j,j+1}$ on the two connected qubits $q_{j+1}$ and $q_{j}$. Finally, we need to bring back the state encoded in qubit $q_{j+1}$ to the $r$-neighbor qubit $q_{j+r}$. This is achieved by applying the inverse sequence of $\mathrm{SWAP}$ gates $S_r^\dagger$. Summarizing, we obtain the quantum circuit identity shown in Fig.\ref{fig: Quantum_circuit}(e), reading
	\begin{align}
		V_{j,j+r} &= S_r^\dagger V_{j,j+1} S_r\notag\\ &=S_r^\dagger\mathrm{CNOT}_{j,j+1}R_Z(2J_1T)\mathrm{CNOT}_{j,j+1}S_r.
	\end{align}

This enables us to realize the desired tunable-range interactions among physically unconnected qubits. However, there is a trade-off involved: the implementation of these interactions requires the insertion of $2(r-1)$ additional $\mathrm{SWAP}$ gates into the quantum circuit. Since each two-qubit gate typically introduces noise, it becomes imperative to optimize our quantum circuit for each Floquet step of the dynamics (as depicted in Fig. \ref{fig: Quantum_circuit}(b)). This optimization involves breaking down each operation into the native gates of the quantum device and strategically reordering our quantum circuit to minimize the involvement of two-qubit gates. A comprehensive description of this procedure is provided in Appendix \ref{app:Quantum circuit optimization}.

Furthermore, we will need to mitigate effect of the noise as the interaction rate increases. We address this problem in the following sections.

 \section{The Role of Noise and Noise Mitigation}
\begin{figure}
	\centering
	\includegraphics[width=\linewidth]{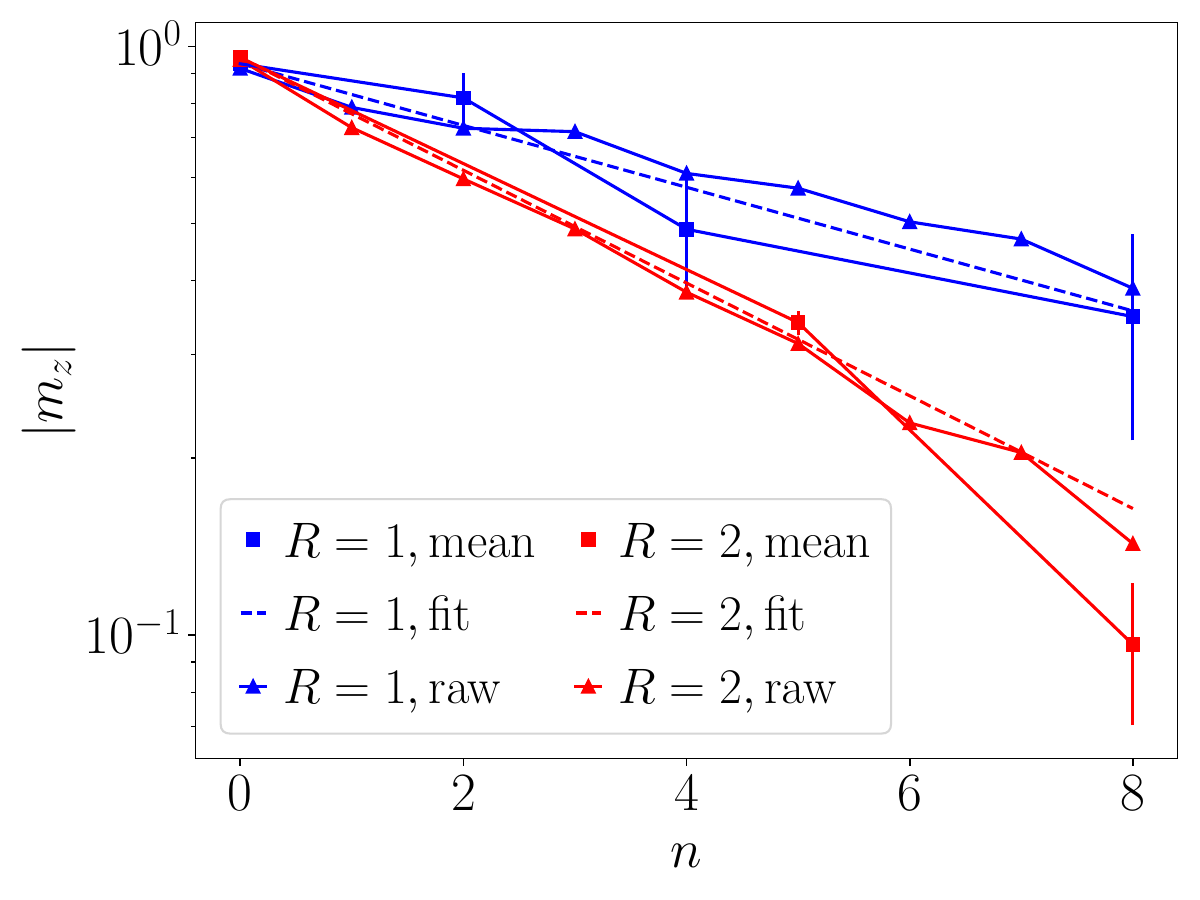}
	\caption{Modulus of the magnetization $|m_z|$ as a function of the stroboscopic times $n$ for nearest neighbor $R = 1$ (blue points) and next-to-nearest neighbor $R = 2$ (red points) Ising interactions. Triangles represent the raw experimental data measured on our quantum simulation of the \textit{ibmq\_mumbai} quantum processor, which involves $N = 18$ qubits undergoing a kicked Ising dynamics with kick angle $\phi = \pi/2+\epsilon$ with $\epsilon = 0.2$. Square points and the corresponding error bars represent the estimators for the average magnetization and its statistical error $\mathbb{E}(m_z)\pm2\sigma(m_z)$, obtained through statistical bootstrapping. Dashed lines represent the best fit of the data with an exponential decay.}
	\label{fig: mz_raw_18}
\end{figure}
The analysis of the raw experimental data clearly demonstrates that the decay of magnetization is predominantly influenced by the effect of noise, as illustrated in Fig. \ref{fig: mz_raw_18}. The figure depicts the absolute value of the average magnetization $|m_z|$ as a function of the stroboscopic time $n$ for nearest neighbor ($R = 1$, in blue) and next-to-nearest neighbor ($R = 2$, in red) interactions. The raw experimental data (triangles in Fig. \ref{fig: mz_raw_18}) are obtained by running $n$ repetitions of the quantum circuit corresponding to a single Floquet step $U_F$, as depicted in Fig. \ref{fig: Quantum_circuit}(b), on the \textit{ibmq\_mumbai} quantum processor using $N = 18$ qubits. At the end of each quantum evolution, a projective measurement of each qubit in the $Z$ basis is performed. To collect sufficient statistics, the experiments for each value of $n$ and $R$ are repeated over a sample of size $\mathcal{N} = 2^{13}$, allowing us to compute the sample average $\langle Z_i \rangle$ over the measurement outcomes. Finally, the spatial average of the magnetization over different sites of the processor is computed as
\begin{align}
	m_z = \frac{1}{N} \sum_{i = 1}^{N} \langle Z_i \rangle,
\end{align}
where $N = 18$ in our case. 

Estimating the statistical error from multiple instances of each quantum simulation to evaluate the sample mean $\mathbb{E}(m_z)$ and standard deviation $\sigma(m_z)$ is not feasible due to the time taken to produce each magnetization estimate. Instead, we rely on the statistical tool of bootstrapping, which is further detailed in Appendix \ref{app:Statistical Bootstrapping}, to generate resampled data from the empirical measurement outcomes. Since the bootstrapped data conforms to the central limit theorem, we may assume normality and evaluate $\mathbb{E}(m_z)$ and $\sigma(m_z)$ from these artificially generated samples of data. The results for the statistical averages $\mathbb{E}(m_z(n))$ are represented by squares in Fig. \ref{fig: mz_raw_18}, while we use two standard deviations, $2\sigma(m_z(n))$, as statistical errors, depicted as error bars in the plots. Notably, we observe that the statistical error increases with the number of Floquet steps involved in the dynamics $n$. This can be understood by a simple statistical argument: we are trying to sample a quantity, the modulus of the magnetization $|m_z|$, which exponentially decreases with $n$. Consequently, the resolution with which we can estimate this quantity deteriorates as $m_z$ approaches the value $m_z\sim e^{-\gamma n}\sim 1/\sqrt{\mathcal{N}}$, i.e., the statistical uncertainty due to the finite size of the sample increases as we approach the stroboscopic time $n\sim (1/2\gamma)\ln\mathcal{N}$.

The decay of magnetization with stroboscopic time $n$ can be described by an exponential fit $|\mathbb{E}(m_z(n))| = ae^{-b n}$, which is obtained using a weighted least squares regression method. This approach accounts for points with high statistical uncertainty, penalizing them in the extrapolation. The resulting exponential decay is depicted as a dashed line in Fig. \ref{fig: mz_raw_18}, showing a rapid decline with increasing $n$. Notably, the decay rate is more pronounced for next-to-nearest neighbor interactions ($R = 2$) compared to nearest neighbor interactions ($R = 1$). This discrepancy can be attributed to the fact that the quantum circuit implementing next-to-nearest neighbor interactions involves more gates, resulting in larger noise effects.

In order to effectively simulate desired physical phenomena in a quantum system, it is crucial to account for and mitigate the detrimental effects of noise. Real-world quantum hardware is susceptible to various sources of errors, such as noisy gates, environmental decoherence, and spurious time dependence of circuit parameters \cite{Ippoliti2021PRXQuantum}. To explicitly model these errors, a common approach is to consider one- and two-qubit depolarizing channels that act on the system's state $\rho$. Specifically, after each single-qubit gate acting on qubit $i$, the single-qubit channel $\phi_i^{1q}$ is applied, while after each two-qubit gate on bond $(i,j)$, the two-qubit channel $\phi_{i,j}^{2q}$ is applied. These channels are defined as \cite{NielsenChuang2010,Ippoliti2021PRXQuantum}

	\begin{align}
		\phi_i^{1q}(\rho) &= (1-p_1)\rho +\frac{p_1}{3}(X_i\rho X_i +Y_i\rho Y_i +Z_i\rho Z_i) \\
		\phi_{i,j}^{2q}(\rho) &= (1-p_2)\rho +\frac{p_2}{15}\sum_{\alpha,\beta = 1}^{3}(\sigma_{\alpha,i}\sigma_{\beta,j}\rho \sigma_{\alpha,i}\sigma_{\alpha,j}),
	\end{align}
	where $\sigma_{1,i} = X_i$, $\sigma_{2,i} = Y_i$, and $\sigma_{3,i} = Z_i$ are the Pauli matrices for qubit $i$, and $\sigma_{\alpha,i}$ and $\sigma_{\beta,j}$ are the corresponding matrices for qubits $i$ and $j$, respectively. By studying the dynamics of the $Z_i$ operators under these depolarizing channels, we can estimate the magnetization decay rate induced by the noisy gates.
	
	To isolate the effect of noise, we consider the case of perfect kick dynamics with $\epsilon = 0$. Under this condition, $Z_i$ is invariant under the two-qubit Ising interaction gates and simply acquires a minus sign under the $\pi$ rotation around the $x$-axis. However, after each two-qubit gate, $Z_i$ decays under $\phi_{i,j}^{2q}$ as
	
	\begin{align}
		\phi_{i,j}^{2q}(Z_i) = (1 -16p_2/15)Z_i,
	\end{align}
	
	and after each single-qubit gate as
	
	\begin{align}
		\phi_i^{1q}(Z_i) = (1 -4p_1/3)Z_i.
	\end{align}
	
	Overall, $Z_i$ decays to $-e^{-\gamma_{\mathrm{dep}}}Z_i$, over one noisy Floquet step with perfect kicks, with $\gamma_{\mathrm{dep}}$ given by
	\begin{align}
		\gamma_{\mathrm{dep},R} = -\ln[(1 -16p_2/15)^{\mathcal{Q}_{2q,R}}(1 -4p_1/3)^{\mathcal{Q}_{1q,R}}],
	\end{align}
	where $\mathcal{Q}_{2q,R}$ and $\mathcal{Q}_{1q,R}$ are the number of two-qubit and single-qubit gates involved in a Floquet step quantum circuit with $R$-neighbor interactions. A naive estimate of these numbers based on the general method previously introduced would yield $\mathcal{Q}_{2q,R} = 2R^3+2$ and $\mathcal{Q}_{1q,R} =  2R+5$, respectively (see Appendix \ref{app:Quantum circuit optimization} for more details). However, for the specific case of the kicked Ising model considered in our quantum simulations, we were able to optimize the quantum circuits corresponding to $R = 1,2$ Floquet steps, reducing the number of two-qubit native gates, involved in the quantum circuit longest path, to $\mathcal{Q}_{2q,R} = 9R^2-14R+7$, with a reduction from cubic to quadratic range dependence. In particular for $R = 1,2$ we have
	\begin{align}
		\mathcal{Q}_{2q,R} =\begin{cases}
			4  &R = 1\\
			15 &R = 2
		\end{cases},\quad
		\mathcal{Q}_{1q,R} = \begin{cases}
			7 &R = 1\\
			9 &R = 2
		\end{cases}.
	\end{align}
	More details on the quantum circuit optimization strategy are provided in Appendix \ref{app:Quantum circuit optimization}.
	
	Another source of noise arises from the finite decoherence time $T_1$ of the qubits, which introduces an additional time scale contributing to the  magnetization decay. Taking into account all the contributions, we can estimate the decay rate of magnetization for a Floquet step with imperfect kicks of an angle $\phi = \pi/2+\epsilon$ to be approximately given by
	\begin{align}
		\Gamma_{1,R} \approx \gamma_{\mathrm{dep},R}+\tau_{R}/T_1 +\gamma_{\epsilon,R},
	\end{align}
	where $\tau_R$ represents the time required to practically implement the Floquet step on the quantum hardware. This can be estimated as
	\begin{align}
		\tau_R = \mathcal{Q}_{1q,R}\tau_{1q}+\mathcal{Q}_{2q,R}\tau_{2q}+\tau_m,
	\end{align}
	where $\tau_{1q}$ and $\tau_{2q}$ denote the time needed to execute each single-qubit and two-qubit gate, respectively, while $\tau_m$ represents the readout time required for measurements. Estimates of these quantities, as obtained from the engine calibration, are provided in the Supplementary Material.
	
	A third source of errors arises from readout errors, which can be modeled as a stochastic process where the outcome of a qubit-state measurement (in the $Z$ computational basis) is randomly flipped with a probability of $p_m$ away from its correct value \cite{Ippoliti2021PRXQuantum}. Specifically, if we define the probability that qubit $i$ points up (down) at time $n$ as $\Pi_{\pm} = \langle (1\pm Z_i(n))/2\rangle$, then the result of the noisy measurement process is $Z_i = \pm 1$, with a probability of $\tilde{\Pi}{\pm}^{(n)} =\Pi{\pm}(1-p_m)+\Pi_{\mp}p_m$. Accordingly, the estimate for the expectation value of $Z_i$ becomes
	\begin{align}
		\tilde{\Pi}_{+}-\tilde{\Pi}_{-}&= (1-2p_m)(\Pi_{+}-\Pi_{-})\notag\\
		&=(1-2p_m)\langle Z_i(n)\rangle.
	\end{align}
	Hence, averaging over positions yields $\tilde{m}_z = (1-2p_m)m_z$, i.e., a damping by a time-independent and range-independent overall prefactor $C_m = (1-2p_m)$.
	
	The inclusion of noise in our model provides a compelling explanation for the rapid exponential decay of magnetization, as observed in Fig. \ref{fig: mz_raw_18}. Moreover, by inserting the estimated values of the parameters $p_1$, $p_2$, $\tau$, and $T_1$, which were extracted from the calibration data provided by IBM and detailed in the SM Appendix, we find that the calculated decay rate is in good agreement with that obtained from fitting the experimental data with a stroboscopic time dependence of the form predicted by our theoretical model,
	\begin{align}
		|m_z(n)| = C_m e^{-n\Gamma_{1,R}}.
	\end{align}
    This understanding of the noise effect justifies our exploration of the possibility of mitigating it through a technique called zero noise extrapolation.
	
	The basic idea of zero noise extrapolation is to intentionally increase the noise level by amplifying the depth of the quantum circuits by a factor of $s$ through a procedure called circuit folding (see Appendix \ref{app: Circuit folding and Zero noise extrapolation} for a more detailed description). Subsequently, we perform quantum simulations for different noise scales $s$, and for each scale, we extract the magnetization decay rate from the measured data. Our noise model then allows us to theoretically estimate the decay rate at noise scale $s$ as
	\begin{align}
		\Gamma_{s,R} \approx s(\gamma_{\mathrm{dep},R}+\tau_{R}/T_1) +\gamma_{\epsilon,R}.\label{eq: Gamma s}
	\end{align}
	Accordingly, a linear fit of the measured decay rates with respect to the parameter $s$ enables us to separate the contribution coming from the noise, $\gamma_\mathrm{noise} = \gamma_{\mathrm{dep},R}+\tau_{R}/T_1$, from $\gamma_{\epsilon,R}$, which represents the decay rate due to the internal system thermalization that destroys the time crystalline order at finite $\epsilon$, and should be stabilized by the presence of longer-range interactions. More precisely, $\gamma_{\epsilon,R}$ is obtained as the zero noise extrapolation of the decay rate in \,\eqref{eq: Gamma s}, $\gamma_{\epsilon,R}\approx \Gamma_{s = 0,R}.$
	\begin{figure}
		\centering
		\includegraphics[width=\linewidth]{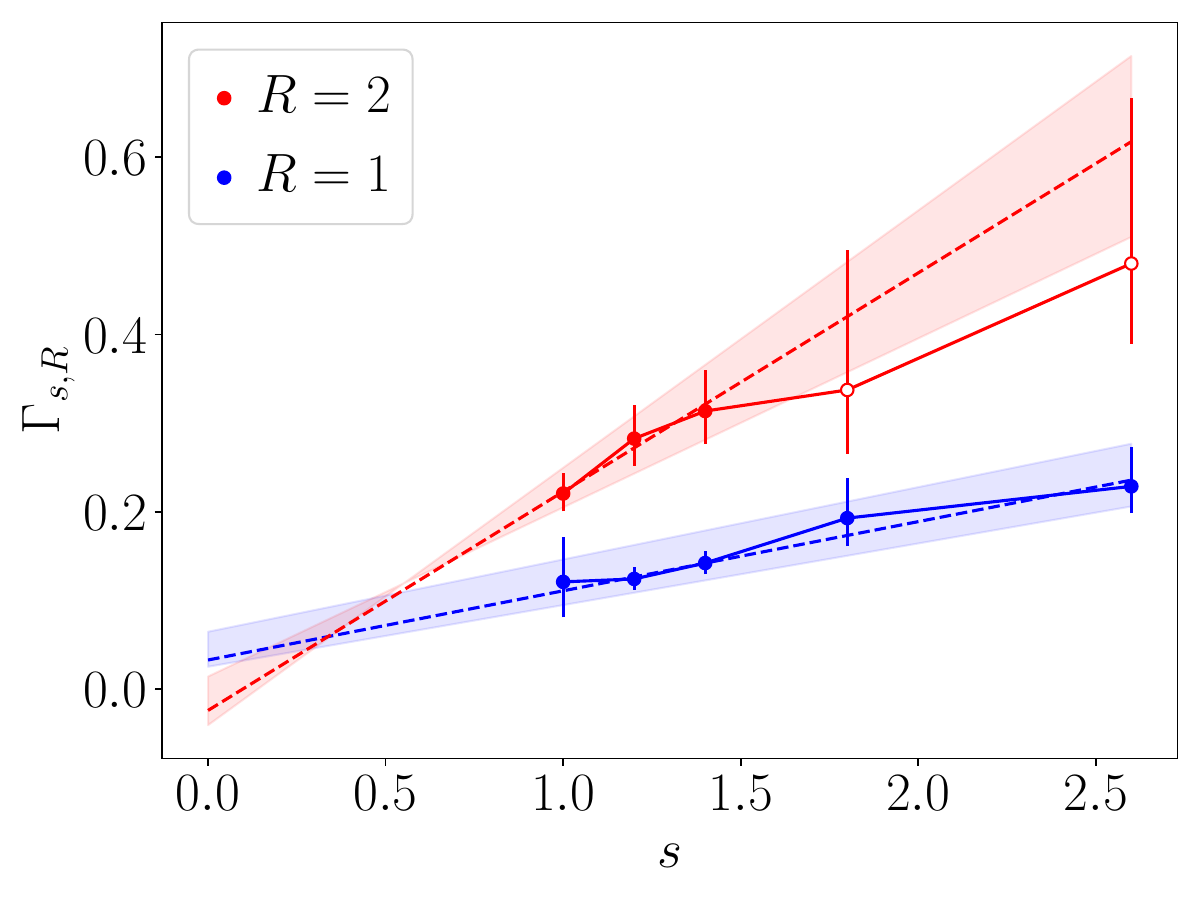}
		\caption{Decay rate of magnetization as a function of noise scale $s$ for $R = 1$ (blue points) and $R = 2$ (red points). Error bars represent two standard deviations $\sigma(m_z)$ estimated through statistical bootstrapping. Dashed lines indicate the best linear fit obtained using weighted least squares regression.  Empty points are excluded from the fitting data.}
		\label{fig: gamma}
	\end{figure}
	The results of this procedure are shown in Fig. \ref{fig: gamma}, where the measured decay rate is plotted as a function of the noise scale $s$. To estimate $\Gamma_{s,R}$ and its uncertainty $\delta\Gamma_{s,R}$, we first estimate the magnetization as a function of the stroboscopic time $n$ at different values of $s$ and $R$ from the measured data, along with the corresponding statistical uncertainty from the standard deviation obtained through the statistical bootstrap method, $\mathbb{E}(m_z)\pm 2\sigma(m_z)$. Then, the decay rate and its uncertainty are obtained through the exponential fit
	\begin{align}
		|\mathbb{E}(m_z)\pm 2\sigma(m_z)| = (C_m\pm\delta C_m)e^{-n(\Gamma_{s,R}\pm\delta \Gamma_{s,R})}.
	\end{align}
	In particular, the exponential fit is performed using weighted least squares regression, and the last two points with $R = 2$ and $s>1.5$ are excluded from the fitting data (empty points in Fig. \ref{fig: gamma}). This exclusion is justified by the fact that the decay rate for these points falls within the range of $0.2<\Gamma_{s>1.5}<0.8$, and thus, the magnetization can be reliably estimated only for stroboscopic times $n<n^*\approx(1/2\gamma)\ln \mathcal{N}$, where $6<n^*<22$. Therefore, not all the time steps $1<n<16$ considered in the exponential fit of $m_z(n)$ from which we extracted this decay rate are within the reach of our statistical resolution. The difficulty of establishing a reliable bootstrap-estimated value confirms this phenomenon, as shown in Fig. \ref{fig: gamma}, where the statistical error bars for these points are significantly larger than those for the other points, indicating the challenge of obtaining a trustworthy value for the magnetization in this regime. Despite the failure of the bootstrap procedure, we include these data as empty points in the plot for completeness, noting that the corresponding error bars are sufficiently large that the resulting fit is still compatible with these unreliable values within $\pm 2\sigma(m_z)$.
	
	Remarkably, upon extrapolation to the zero noise limit, the decay rate of the $R = 2$ case is found to be smaller than that in the nearest neighbors case $R = 1$. Specifically, we obtain
	\begin{align}
		\Gamma_{0,2}\pm \delta \Gamma_{0,2}<\Gamma_{0,1}\pm \delta \Gamma_{0,1}.
	\end{align}
    Most significantly, we find that the decay rate analytically predicted from
the theoretical model, $\gamma_{\epsilon,R}\approx\epsilon^{2R+1}$, is compatible with the
extrapolated values within the estimated
uncertainty, i.e., 
\begin{align}
    \gamma_{\epsilon,R}\in[\Gamma_{0,R}-\delta\Gamma_{0,R},\Gamma_{0,R}+\delta\Gamma_{0,R}],
\end{align}
indicating that the extrapolated decay rate is consistent with the theoretical expectations within the statistical uncertainty $\delta \Gamma_{0,R}$, which has been estimated by extrapolating $\delta \Gamma_{s,R}$ to $s = 0$.

 \section{Conclusion}
	In this paper we have demonstrated the potential of superconducting quantum hardware for advancing the field of digital quantum simulation by investigating the possibility of implementing quantum dynamics of systems with couplings beyond nearest neighbors. We have utilized the universality of the native gates in quantum processors to implement couplings among physically disconnected qubits and carefully mitigated the effects of gate noise, measurement errors, and statistical errors from the raw results. Our focus has been on the stabilization of discrete Floquet time-crystalline response as the interaction range increases, and we have implemented,  on IBM quantum superconducting processors, a quantum simulation of a Floquet-driven quantum spin chain with interactions beyond nearest neighbors.
	
	Our results, as shown in Fig.\,\ref{fig: mz_raw_18}, reveal that the magnetization dynamics under a kicked Floquet driving exhibits a fast exponential decay dominated by noise in the raw data, with a faster decay rate for larger interaction ranges due to the increased depth of the quantum circuit. However, after applying the zero noise extrapolation procedure, we were able to separate the role of noise from the true decay caused by the dynamical generation of excitations in the system during the Floquet driving. The mitigated data for the magnetization decay rate, in Fig.\,\ref{fig: gamma}, show a clear trend of slower decay in the zero noise limit compared to the raw data, indicating the effectiveness of our error mitigation approach.
	
	Furthermore, we have estimated the statistical uncertainty of the mitigated data, represented by the error bars in Fig. \ref{fig: gamma}, which grows with the noise level as expected. Importantly, we have observed that the regions corresponding to the zero noise extrapolated values of the magnetization decay rate for different interaction ranges do not intersect within the considered values of the parameters, indicating that the effects of noise have been effectively removed from the data.
	
	We have also compared our results with the theoretical expectations, and found that the magnetization decay rate analytically predicted from the theoretical model, $\gamma_{\epsilon,R} \approx \epsilon^{2R+1}$, is compatible with the final results of our quantum simulation within the estimated uncertainty. This agreement between theory and experiment provides evidence for the validity of our approach in simulating quantum systems with interaction ranges beyond the limits imposed by hardware connectivity. 
	
	It is important to highlight that,in principle, the quantum circuit we used for implementing beyond nearest-neighbor interactions on quantum superconducting computers has the potential to enable tunable interaction ranges. However, there is a fundamental limitation that led us to restrict our quantum simulations to cases where $R=1,2$. This is due to the fact that increasing the effective range of interaction will inevitably require a deeper quantum circuit, which in turn increases the impact of noise. Practically speaking, we anticipate that already at $R = 3,4$, the noise level may be significant enough to prevent a reliable estimate of the magnetization decay. Another way of looking at this is that for the result to be within our resolution, the noise level should be at most comparable with the decay rate $\gamma_{\epsilon, R} \approx \epsilon^{2R+1}$. For $R \geq 3$, this noise level could already exceed the threshold for fault-tolerant quantum error correction and therefore surpass the capabilities of the available NISQ devices.
	
	Nevertheless, there are several steps that can be taken to overcome this problem. For example, the noise level of quantum devices is expected to decrease as the performance of available quantum computers improves. Additionally, classical algorithms can be used to further optimize the quantum circuit we proposed, reducing the number of two-qubit gates involved. Furthermore, it would be instructive to benchmark our results on different experimental platforms that naturally allow for the implementation of long-range interactions, such as trapped ions or Rydberg atoms devices. These exciting problems are beyond the scope of the present work and hence we leave them for future research.
	
	Another crucial aspect to emphasize is that in our quantum simulation, we utilize 18 functional qubits out of the total 27 nominal qubits of the \textit{ibmq\_mumbai} device. This distinction is significant since the number of functional qubits is often smaller in other applications. For instance, in quantum chemistry applications, as described in Ref. \cite{Weaving2023arXiv}, noise mitigation strategies are employed to simulate molecules whose Hamiltonian can be encoded in only three qubits in order to achieve the desired chemical accuracy. This exemplifies the suitability of Floquet dynamics, similar to those analyzed in our work, for implementation on NISQ quantum computers.
	
	In conclusion, our quantum simulation on IBM quantum superconducting processors has demonstrated the potential of these platforms for simulating quantum systems with couplings beyond nearest neighbors and has offered insights into the fundamental physics of long-range systems. Our error mitigation approach has been effective in removing the effects of noise and measurement errors from the raw results, and the mitigated data are in good agreement with theoretical expectations. This work opens up new possibilities for studying quantum systems with long-range interactions and paves the way for further advancements in the field of digital quantum simulation on superconducting quantum hardware.

\begin{acknowledgments}
 N.D. acknowledges funding by the Swiss National Science Foundation (SNSF) under project funding ID: 200021 207537 and by the Deutsche Forschungsgemeinschaft (DFG, German Research Foundation) under Germany’s Excellence Strategy EXC2181/1-390900948 (the Heidelberg STRUCTURES Excellence Cluster). This work is part of the MIUR-PRIN2017 project Coarse-grained description for nonequilibrium systems and transport phenomena (CO-NEST) No. 201798CZL. AS and SS acknowledge acknowledge financial support from National Centre for HPC, Big Data and Quantum Computing (Spoke 10, CN00000013). Access to the IBM Quantum Computers was obtained through the IBM Quantum Hub at CERN with which the Italian Institute of Technology (IIT) is affiliated.
\end{acknowledgments}

\appendix
\section{Quantum circuit optimization}
\label{app:Quantum circuit optimization}
	\begin{figure*}
		\centering
		\includegraphics[width=\linewidth]{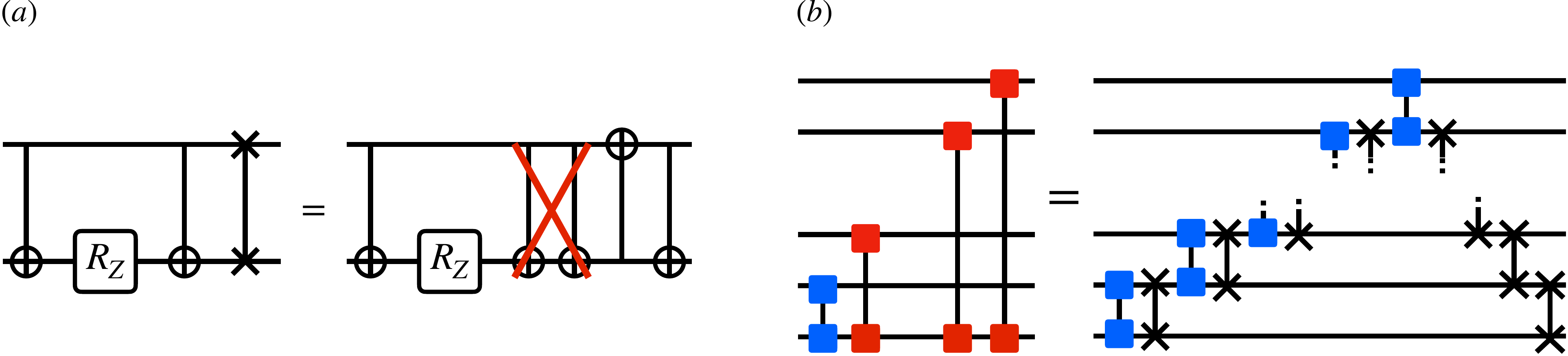}
		\caption{Quantum circuit optimization techniques using circuit identities. (a) Cancellation of $\mathrm{CNOT}$ gates in adjacent $V_{j,j+1}$ and $\mathrm{SWAP}$ gates. (b) Rearrangement of the quantum circuit to implement Ising interactions with $r = 1,\dots, R$ ranges while maximizing the number of adjacent $V_{j,j+1}$ and $\mathrm{SWAP}$ gates.}
		\label{fig: Cricuit_optimization}
	\end{figure*}
 As discussed in the main text, minimizing the number of operations in the quantum circuit for implementing the Floquet dynamics is crucial due to the increase in noise with each quantum gate, resulting in a rapid magnetization decay. In particular, two-qubit gates are more prone to errors, so our focus is on reducing their number in our circuits.
		
	We start by obtaining an estimate of the number of operations required to implement the quantum circuit shown in Fig. 2b of the main text using the native gates of the \textit{ibmq\_mumbai} processor used in this work, without any optimization. The native gates include the controlled-$\mathrm{NOT}$ gate ($\mathrm{CNOT}$), identity gate ($\mathrm{ID}$), rotations along the z-axis ($R_Z$), the $\mathrm{NOT}$ gate ($X$), and the $\mathrm{SX} = \sqrt{X}$ gate.
		
	Regarding single-qubit gates, only the rotations around the $X$ axis, corresponding to the Floquet driving kicks, need to be further decomposed into native gates, which can be efficiently done as follows
	\begin{align}
		R_X(\phi) = R_Z(\pi/2)\sqrt{X}R_Z(\phi)\sqrt{X}R_Z(5\pi/2).
	\end{align}
	Thus, each kick requires five additional single-qubit gates, resulting in a total of $Q_{1q,R} = 2R+5$ gates.
	The only native two-qubit gate available is the $\mathrm{CNOT}$ gate. To estimate $Q_{2q,R}$, we need to count the number of $\mathrm{CNOT}$ gates involved in the hardware implementation of each Floquet step. As shown in Fig. 2c of the main text, each nearest-neighbor Ising interaction is implemented using two $\mathrm{CNOT}$ gates. Moreover, each $\mathrm{SWAP}$ gate is realized using three $\mathrm{CNOT}$ gates, as it can be decomposed as
	\begin{align}
		\mathrm{SWAP}_{j,j+1} = \mathrm{CNOT}_{j,j+1}\mathrm{CNOT}_{j+1,j}\mathrm{CNOT}_{j,j+1}.
	\end{align}
	Moreover, each $r$-range interaction is implemented by adding $2(r-1)$ $\mathrm{SWAP}$ gates to the nearest-neighbor interaction. Therefore, each $r$-range interaction gate requires $2+6(r-1)$ $\mathrm{CNOT}$ gates for implementation. If we want to realize interactions with ranges $r = 1, \ldots, R$, then the longest path, determining the circuit depth, contains $r$ non-parallelizable copies of each $r$-range operation for $r>1$ and two copies for $r = 1$. Summing up all the contributions, we obtain:
	\begin{align}
		Q_{2q,R} = \sum_{r>1}^R(2+6r(r-1))+4 = 2R^3+2.
	\end{align}
		
	We can optimize the structure of our quantum circuit (Fig. 2(b) of the main text) to reduce its depth and complexity. First of all, we notice that, as shown in Fig. \ref{fig: Cricuit_optimization}(a), each time we have a sequence $V_{j,j+1}\mathrm{SWAP}_{j,j+l}$, we can use the fact that $\mathrm{CNOT}^2 = \mathbb{I}$ to eliminate two adjacent $\mathrm{CNOT}$ gates. To systematically exploit this fact, we can rearrange our quantum circuit using the circuit identity in Fig. \ref{fig: Cricuit_optimization}(b). Here, we utilize the properties $[V_{j,j+l},V_{j,j+r}] = 0$ for all $l,r$, and $[V_{j,j+1},\mathrm{SWAP}_{j,j+l}] = 0$ to maximize the number of adjacent $V_{j,j+1}\mathrm{SWAP}_{j,j+l}$, and thereby increase the number of $\mathrm{CNOT}$ gates that cancel out. 
	for a circuit implementing a sequence of Ising interactions of ranges $r = 1,\dots R$, we can cancel up to $2(R-1)$ $\mathrm{CNOT}$ gates using this trick. The depth of each subcircuit of this form is then given by
	\begin{align}
		2R+6(R-1)-2(R-1) = 6R-4.
	\end{align}

	To realize a kicked Ising model with interaction range up to $R$ in a chain of $N$ qubits, we can divide the $N$ qubits into subsets of size $2R$ that can be processed in parallel. To compute the circuit depth, which refers to the number of operations in the longest path, we can focus on only one subset at a time. Each subset contains $R$ subcircuits with interaction ranges $r = 1, \ldots, R$, following the form shown in Fig. \ref{fig: Cricuit_optimization}(b). The depth of each of these subcircuits is $6R-4$. The remaining $R$ subcircuits include interactions of range $r = 1, \ldots, l$, where $l$ varies from $l = 1$ to $l = R-1$, corresponding to a depth of $6l-4$ for each circuit. By summing up all the contributions, we obtain the optimized number of $\mathrm{CNOT}$ gates as:
	
	\begin{align}
		Q_{2q,R} &= R(6R-4) + \sum_{l =1}^{R-1}(6l-4)\notag\\
		&= 9R^2-11R+4.
	\end{align}
	Finally, we observe that the last sequence of $\mathrm{SWAP}$ gates in the circuit shown in Fig. \ref{fig: Cricuit_optimization}b is only necessary if we need to apply different gates on different qubits after that. If this is not the case, we can simply substitute the $\mathrm{SWAP}$ gates with a relabeling of the qubit numbers, which must be taken into account when reading the final measurement outcomes. This fact allows us to eliminate $(R-1)$ $\mathrm{SWAP}$ gates and $3(R-1)$ $\mathrm{CNOT}$ gates in the last subcircuit of this form. Therefore, we obtain:
	
	\begin{align}
		Q_{2q,R} = 9R^2-14R+7.
	\end{align}
	As a final remark, we note that for large values of maximum range $R$, and hence large circuit complexity, additional simplifications of the circuit may be possible by using optimized relabelings of the qubit numbers during the evolution, which can increase the number of parallelizable operations. However, such an optimization strategy is circuit and range dependent, and can only be carried out numerically or in an approximate manner. On the other hand, for the case of $R = 2$ that we considered in our quantum simulation, we can claim that our circuit is optimal with respect to the number of $\mathrm{CNOT}$ gates involved.
	\section{Circuit folding and Zero noise extrapolation}
    \label{app: Circuit folding and Zero noise extrapolation}
	Zero noise extrapolation (ZNE) is a well-studied error mitigation method in the literature \cite{Li2017PRX, Temme2017PRL, Kandala2019Nature,Weaving2023arXiv}. It is a powerful technique that allows for the estimation of noiseless expectation values of observables from a series of measurements obtained at different levels of noise. The ZNE process involves two steps: intentional scaling of noise and extrapolation to the noiseless limit. In the first step, the target circuit is executed at varying error rates denoted by $s$, with expectation values estimated for the original circuit ($s = 1$) as well as circuits at increased error rates ($s > 1$). Then, in the second step, a function, motivated by physical arguments, is fitted to these expectation values and used to extrapolate to error rate $s = 0$, providing an error-mitigated estimate.
	
	There are various methods to increase the error rate $s$. Examples in the literature include pulse stretching \cite{Temme2017PRL} or, at a gate level, unitary folding \cite{Li2017PRX,Tiron2020IEEE}. In our implementation of ZNE, we increase $s$ using a local unitary folding technique. This technique involves increasing the number of operations by applying a mapping $U \to UU^\dagger U$ to individual gates of the circuit. Specifically, the unitary gates to be folded are randomly chosen from the set of gates composing the circuit in such a way that the circuit depth is approximately increased by the desired factor $s$. This random selection helps to ensure that the circuit is exposed to a variety of gate sequences and interactions, allowing for a more comprehensive study of the circuit's behavior under different noise conditions.
    \section{Statistical Bootstrapping}
    \label{app:Statistical Bootstrapping}
    We utilize the statistical technique of bootstrapping to quantify the uncertainty in our magnetization estimates. In an ideal scenario we would repeat quantum experiments multiple times to obtain a comprehensive understanding of the "true" magnetization distribution.  However, this approach is impractical due to the significant time required for each magnetization estimate. Instead, we conduct the experiment once and generate resampled measurement data from the empirical distribution using bootstrapping, a widely used statistical technique. This method makes the statistical analysis very convenient, and it is then becoming a common practice to estimate the statistical errors in digital quantum simulations \cite{Weaving2023arXiv}.
	
	Let us assume we perform an $\mathcal{N}$-shot quantum experiment and obtain a collection of $\mathcal{N}$ outcomes. Each measurement outcome is represented as a string of 0s and 1s, denoted as $Z_{1,a}\dots Z_{N,a}$, where $Z_i =0,1$, $N$ is the number of measured qubits, and the index $a$ labels the different outcomes ($a = 1,\dots,\mathcal{N}$). The magnetization associated with each string can be computed by averaging over the qubits as follows:
	\begin{align}
		m_{z,a} = \frac{1}{N}\sum_{i = 1}^N Z_{i,a}.
	\end{align}
	This gives us the set of magnetization values ${m_{z,a}}$ with $a = 1,\dots,\mathcal{N}$. We define the empirical magnetization distribution $P_{1}(m_{z,a})$ as the histogram of the ${m_{z,a}}$ set. The average over this empirical distribution, denoted as $m_z$, corresponds to the experimentally obtained quantum expectation value on the final state of the system and can be expressed as
	\begin{align}
		m_z = \frac{1}{N}\sum_{i = 1}^N \langle Z_{i} \rangle =\sum_{a=1}^{\mathcal{N}}P_{1}(m_{z,a})m_{z,a}.
	\end{align}
	\begin{figure}
		\centering
		\includegraphics[width=\linewidth]{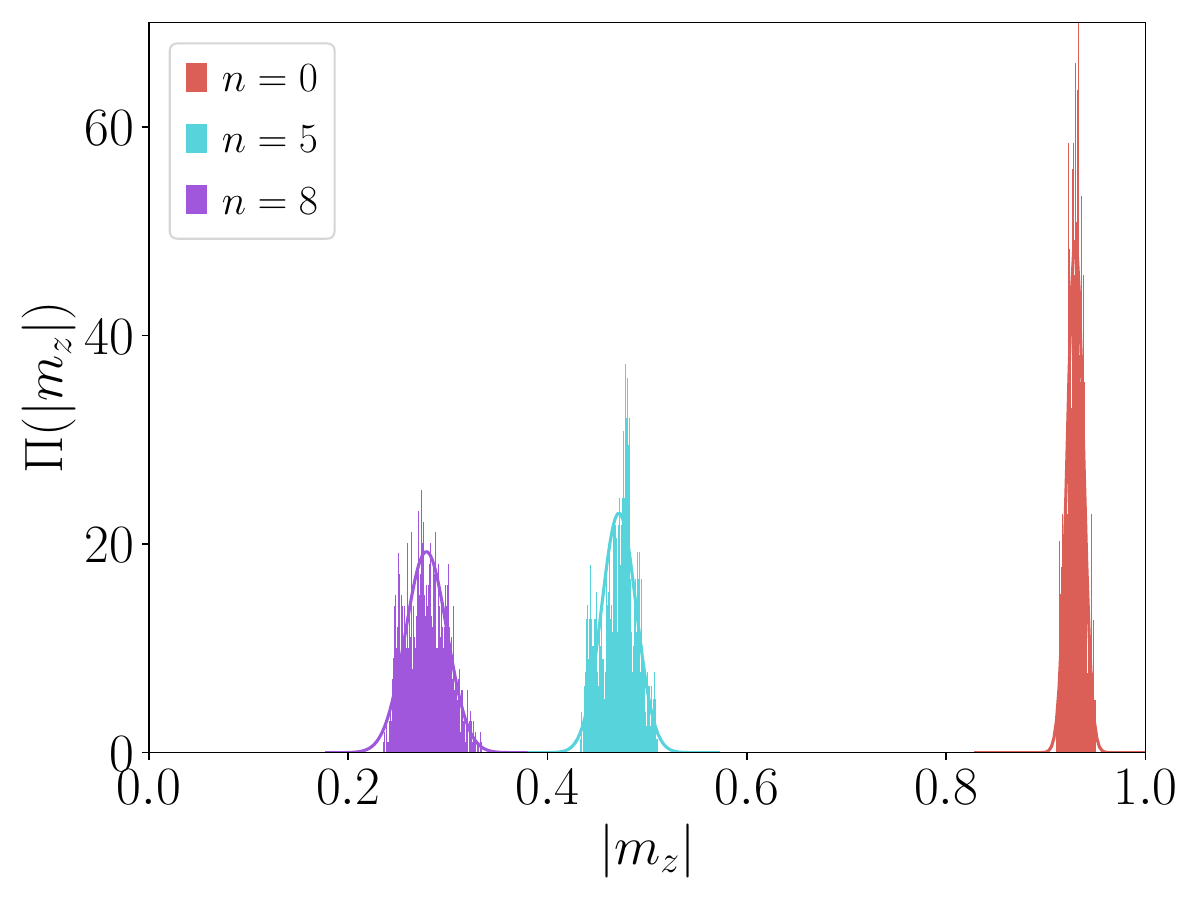}
		\caption{Bootstrap distribution of averages obtained by resampling $M = 1000$ times the measured data of three quantum simulations with range $R = 1$, noise scale $s = 1.4$, and different numbers of Floquet steps $n = 0,5,8$. The number of bins considered in each histogram is $100$. }
		\label{fig: bootstrap_example}
	\end{figure}
	The bootstrapping approach involves resampling from the empirical measurement distribution $P_{1}(m_{z,a})$.  We sample elements from the set ${m_{z,a}}$ (or, equivalently, from the set of strings $\{Z_{1,a}\dots Z_{N,a}\}$) $\mathcal{N}$ times to create a new set of measurement outcomes, and from this, a new empirical distribution $P_{2}(m_{z,a})$.We repeat this process as many times as possible given the available computational resources, say $M$ repetitions, to obtain a set of distributions ${P_{1}, P_{2}, \dots, P_{M}}$.  From each of these distributions, we can compute the average $m_z^{\alpha}$ with $\alpha = 1,\dots, M$, and from the histogram of the set of averages, we obtain their distribution $\Pi(m_z^{\alpha})$.Since each resampling is independent, the distribution of averages should tend to a Gaussian in the large $M$ limit, according to the central limit theorem. Accordingly, we can define our estimator for $m_z$ and its statistical error as the average of the $\Pi(m_z^{\alpha})$ distribution,
	\begin{align}
		\mathbb{E}(m_z) = \sum_{\alpha = 1}^{M}\Pi(m_z^{\alpha})m_z^{\alpha},
	\end{align}
	and its standard deviation,
	\begin{align}
		\sigma(m_z) =\sqrt{\sum_{\alpha = 1}^{M}\Pi(m_z^{\alpha}) (m_z^{\alpha}-\mathbb{E}(m_z))^2}.
	\end{align}
	 This method enables us to obtain error bars in Figs. \ref{fig: mz_raw_18} and \ref{fig: gamma} of the main text as $\mathbb{E}(m_z)\pm2\sigma(m_z)$. Figure \ref{fig: bootstrap_example} shows, as an example, the distributions $\Pi(m_z^{\alpha})$ obtained through $M = 1000$ resamples of the measured data of three quantum simulations with range $R = 1$, noise scale $s = 1.4$, and number of Floquet steps $n = 0,5,8$, respectively. These are compared with Gaussian distributions with the same mean and standard deviation, finding good agreement. We notice that, as expected, the mean $\mathcal{E}(m_z)$ is smaller for a larger number of Floquet steps $n$, signaling the magnetization exponential decay. Moreover, distributions at later stroboscopic times become broader, signaling the growth of the statistical error due to the fact that we are trying to sample a quantity which is exponentially decaying with $n$.
\newpage
\begin{widetext}
\begin{center}
    \textbf{\Large{Supplementary Materials}}
\end{center}

\section{Calibration details of the quantum processor}
	In this section, we provide all the calibration details of the \textit{ibmq\_mumbai} backend, which is one of the 27-qubit IBM Falcon processors used in our experiments. Figure \ref{fig: ibmq_mumbai} shows the topology of the processor and the qubit labeling numbers. We present the corresponding coupling map for two-qubit gates in Figure \ref{fig: CNOT_error}, which includes the $\mathrm{CNOT}$ gate error $p_2$ (panel a) and the $\mathrm{CNOT}$ gate length $\tau_{2q}$ (panel b) for each pair of physically connected qubits on the device. Table \ref{tab:hardware_properties_2} reports the following properties for each qubit of the processor: relaxation time $T_1$, decoherence time $T_2$, single-qubit gate length $\tau_{1q}$, single-qubit gate error $p_1$, readout length $\tau_m$, and readout error $p_m$.
\begin{figure*}[h!]
	\centering
	\includegraphics[width=0.7\linewidth]{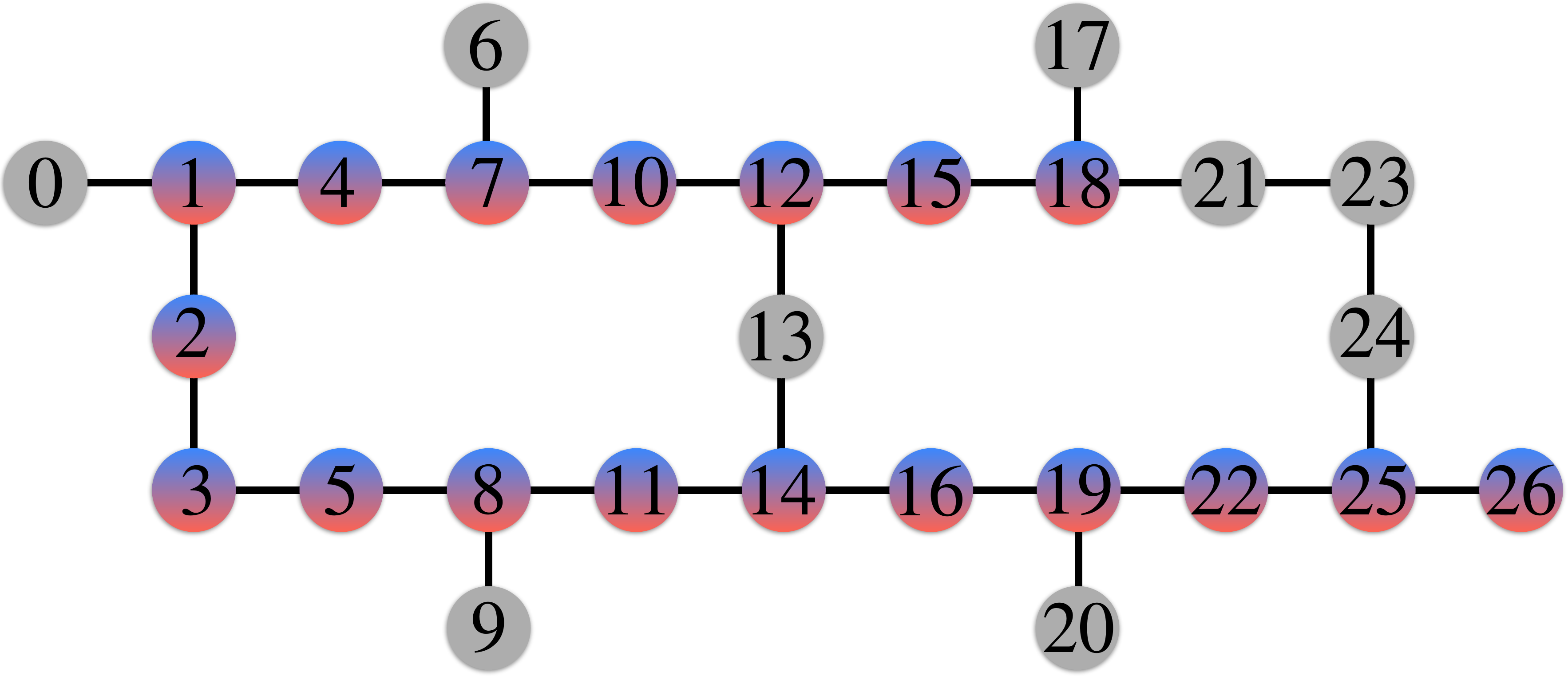}
	\caption{Topology of the \textit{ibmq\_mumbai} quantum processor with numbered qubits. Grey qubits were not involved in our quantum simulations.}
	\label{fig: ibmq_mumbai}
\end{figure*}
\begin{figure*}[h!]
	\centering
	\includegraphics[width=\linewidth]{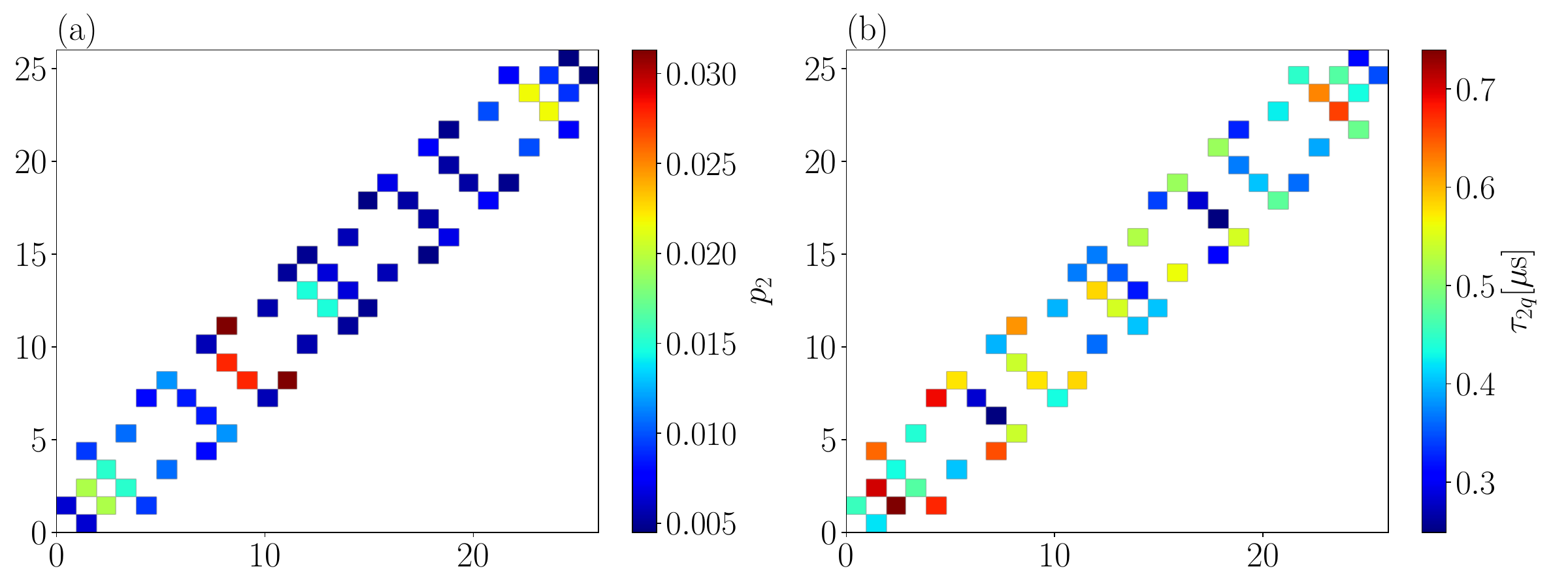}
	\caption{Coupling map and calibration data for the $\mathrm{CNOT}$ gates of \textit{ibmq\_mumbai} quantum processor at the time our experiments were performed. (a) $\mathrm{CNOT}$ error. (b) $\mathrm{CNOT}$ lenght.}
	\label{fig: CNOT_error}
\end{figure*}
\begin{center}
	\begin{table*}[h!]
 \centering
		\begin{tabular}{|p{1.5cm}|p{1.5cm}|p{1.5cm}|p{1.5cm}|p{1.5cm}|p{1.5cm}|p{1.5cm}|}\hline Qubit &$T_1[\mu\mathrm{s}]$ &$T_2[\mu\mathrm{s}]$  &$\tau_{1q}[\mathrm{ns}]$ &$p_1\times 10^4$ &$\tau_m[\mu\mathrm{s}]$ &$p_m\times 10^2$ \\\hline
		$q_{1}$ &139.25 &84.63 &35.56 &2.23 &3.55 &0.69 \\\hline
		$q_{2}$ &79.93 &95.46 &35.56 &8.14 &3.55 &9.65 \\\hline
		$q_{3}$ &129.79 &269.09 &35.56 &2.02 &3.55 &2.19 \\\hline
		$q_{4}$ &108.52 &35.25 &35.56 &2.24 &3.55 &1.77 \\\hline
		$q_{5}$ &96.90 &144.35 &35.56 &1.99 &3.55 &1.20 \\\hline
		$q_{7}$ &104.21 &87.53 &35.56 &1.75 &3.55 &1.60 \\\hline
		$q_{8}$ &122.01 &58.38 &35.56 &3.60 &3.55 &3.69 \\\hline
		$q_{10}$ &140.58 &174.33 &35.56 &1.98 &3.55 &2.25 \\\hline
		$q_{11}$ &217.71 &74.63 &35.56 &2.04 &3.55 &2.62 \\\hline
		$q_{12}$ &216.04 &310.34 &35.56 &1.68 &3.55 &3.73 \\\hline
		$q_{14}$ &162.60 &244.97 &35.56 &1.88 &3.55 &1.48 \\\hline
		$q_{15}$ &70.33 &159.80 &35.56 &1.92 &3.55 &2.44 \\\hline
		$q_{16}$ &100.41 &136.65 &35.56 &1.85 &3.55 &2.61 \\\hline
		$q_{18}$ &55.97 &298.32 &35.56 &1.84 &3.55 &5.66 \\\hline
		$q_{19}$ &164.04 &163.14 &35.56 &1.75 &3.55 &1.36 \\\hline
		$q_{22}$ &146.29 &97.10 &35.56 &1.87 &3.55 &1.31 \\\hline
		$q_{25}$ &185.26 &64.45 &35.56 &1.88 &3.55 &1.67 \\\hline
		$q_{26}$ &67.10 &115.73 &35.56 &2.40 &3.55 &1.82 \\\hline
		\end{tabular}
		\caption{Calibration data for each qubit of the processor of the \textit{ibmq\_mumbai} quantum processor at the time our experiments were performed.}
		\label{tab:hardware_properties_2}
	\end{table*}
\end{center}
\newpage
\subsection{Statistical bootstrap data}
In this section we provide the complete dataset produced by applying the bootstrap procedure to the measured outputs of each quantum simulation performed on the \textit{ibmq\_mumbai} processor in this study. In particular Fig. \ref{fig: bootstrap_R1} shows the bootstrap data for nearest neighbor Ising interactions $R = 1$, different noise scales $s = 1,1.2,1.4,2.6$ and different stroboscopic times $n = 0,5,8,10,12$, while Fig. \ref{fig: bootstrap_R2} shows the same the bootstrap data for next to nearest neighbor Ising interactions $R = 2$, different noise scales $s = 1,1.2,1.4,2.6$ and different stroboscopic times $n = 0,5,8$.
	\begin{figure*}[h!]
		\centering
		\includegraphics[width=\linewidth]{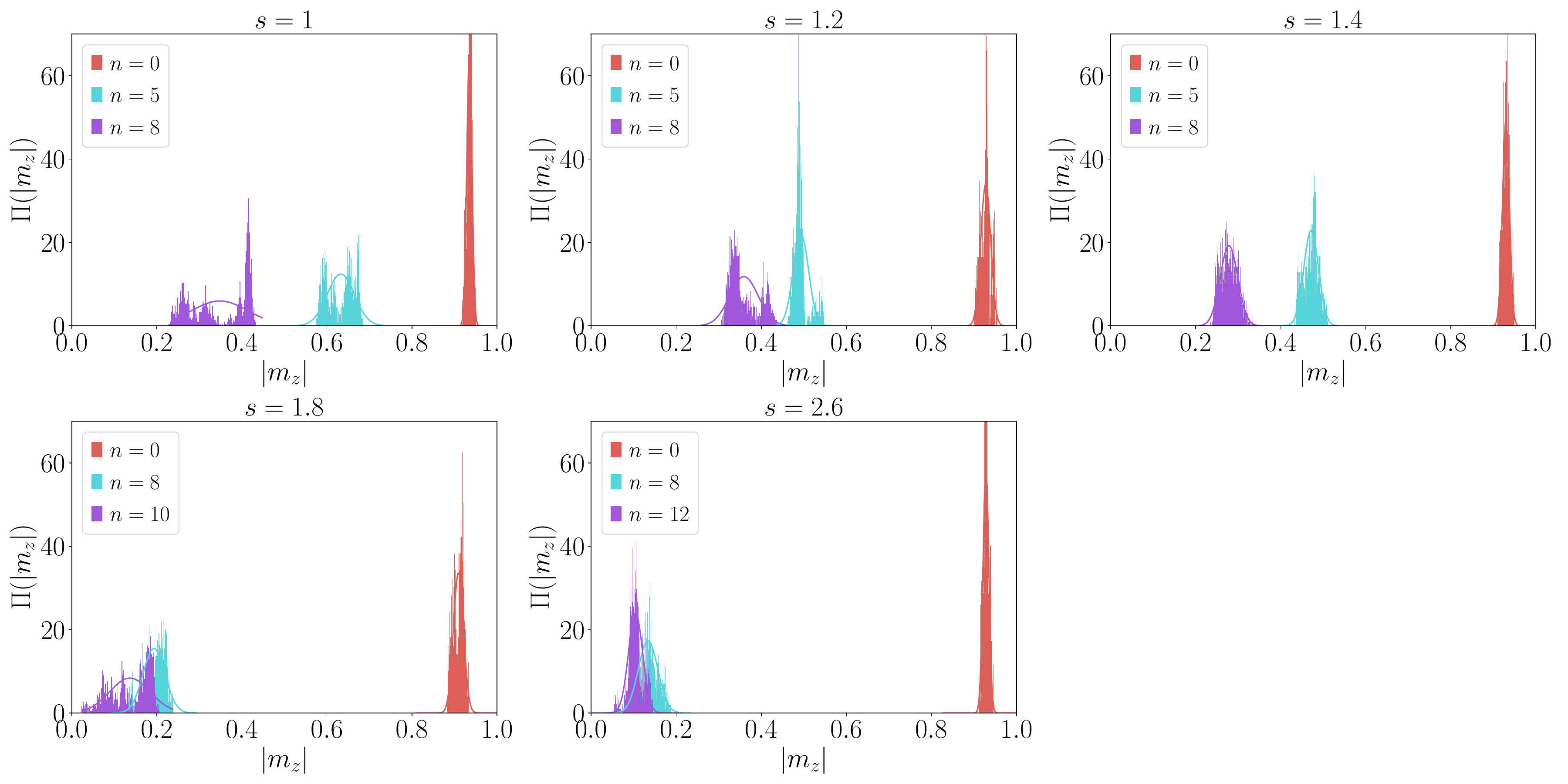}
		\caption{Bootstrap distribution of averages obtained by resampling $M = 1000$ times the measured data of the quantum simulations with range $R = 1$, noise scales $s = 1,1.2,1.4,1.8,2.6$ and different numbers of Floquet steps $n = 0,5,8,10,12$. The number of bins considered in each histogram is $100$. }
		\label{fig: bootstrap_R1}
	\end{figure*}
	\begin{figure*}[h!]
		\centering
		\includegraphics[width=\linewidth]{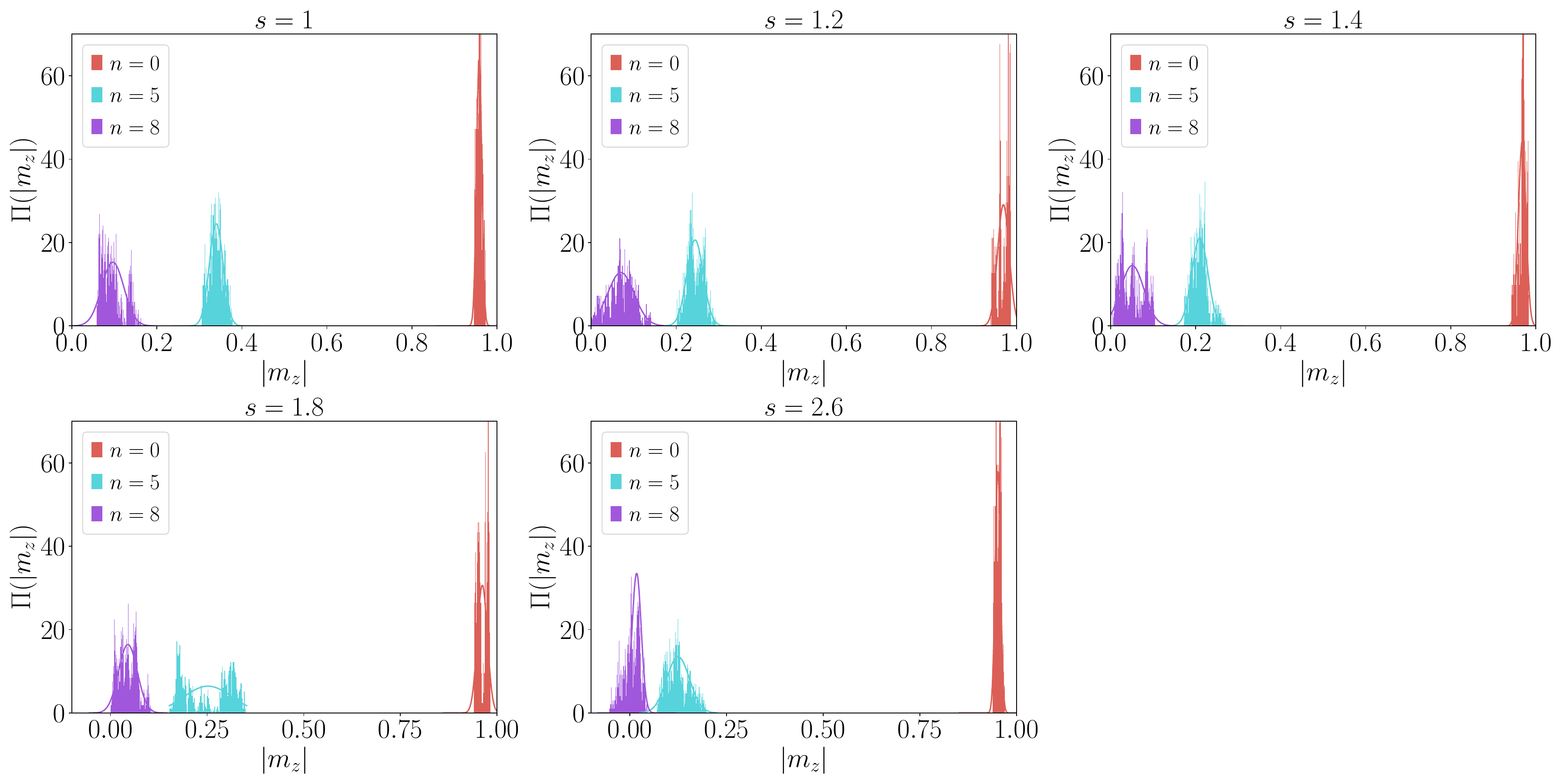}
		\caption{Bootstrap distribution of averages obtained by resampling $M = 1000$ times the measured data of the quantum simulations with range $R = 2$, noise scales $s = 1,1.2,1.4,1.8,2.6$ and different numbers of Floquet steps $n = 0,5,8$. The number of bins considered in each histogram is $100$. }
		\label{fig: bootstrap_R2}
	\end{figure*} 
\end{widetext}
 \newpage
	%
    

%
%

%



\end{document}